\documentclass{iopjournal}

\usepackage{graphicx}%% 多加的
\usepackage{bm}%% 多加的
\usepackage{ragged2e}%%% 多加的
\usepackage{lipsum}%%% 多加的
\usepackage{amsmath}%%% 多加的
\usepackage{amsthm}
\usepackage{mathrsfs}
\usepackage{amsfonts}
\usepackage{amssymb}
\usepackage{textcomp}
\usepackage{appendix}
\usepackage{cite}
\begin{document}

\articletype{Article type} %	 e.g. Paper, Letter, Topical Review...

\title{Scalable long-distance quantum fan-out gates with reduced resource overhead}

\author{Wen-Qiang Liu$^{1,*}$\orcid{0000-0001-9804-2716} and Hai-Rui Wei$^{2,*}$\orcid{0000-0001-7459-4161}}

\affil{$^1$Department of Mathematics and Physics, Shijiazhuang Tiedao University, Shijiazhuang 050043, China}

\affil{$^2$School of Mathematics and Physics, University of Science and Technology Beijing, Beijing 100083, China}

\affil{$^*$Author to whom any correspondence should be addressed.}

\email{wqliu@stdu.edu.cn and hrwei@ustb.edu.cn}

\keywords{quantum gate teleportation, long-distance quantum gate, quantum circuit, fan-out gate, linear optics}

\begin{abstract}
\justifying
Scalable long-distance quantum gates are essential for distributed quantum computation and scalable quantum networks. We propose an efficient scheme for implementing an $n$-qubit fan-out gate $C^n$ (also known as a controlled-NOT-$\cdots$-NOT gate) across spatially separated parties.  Our protocol requires only a pre-shared $n$-qubit Greenberger-Horne-Zeilinger state, $n$ local controlled-NOT (CNOT) gates, and feed-forward operations, thereby reducing both the CNOT and Hadamard gate counts by $(n-2)$ when compared with the standard decomposition-based approach.
We further develop two optical architectures to implement the heralded and post-selected  $n$-photon fan-out gates, respectively.  These schemes require fewer single photons resources and achieve higher success probabilities than their decomposition-based counterparts, representing significant improvements.
The results indicate that our approach is experimentally feasible and pave the way toward  practical long-distance quantum computation.
\end{abstract}

\section{Introduction}  \label{sec1}

\justifying

The computational advantage of quantum computer over its classical counterparts in solving certain intractable problems has been both theoretically established \cite{hangleiter2023computational,liu2024efficiently} and experimentally demonstrated \cite{zhong2020quantum,gao2025establishing}. Prominent examples include Shor's algorithm for integer factorization and Grover's algorithm for unstructured search.
Nevertheless, the ability to store and manipulate a large number of qubits in a single quantum processor, while maintaining high-fidelity control and strong inter-qubit connectivity, remains a formidable challenge with current technology. This limitation stems from persistent physical and technical obstacles, such as decoherence, dissipation, and experimental imperfections \cite{ladd2010quantum,de2021materials,gill2025quantum,dirnegger2026distilled}.
Fortunately, distributed quantum computing (DQC) offers a promising solution by leveraging collaborative computation across multiple small-scale quantum nodes \cite{cirac1999distributed,monroe2014large,barral2025review}. In this modular framework, each node operates independently as either a sender or receiver via classical and quantum channels. Such an architecture allows for the collective execution of large-scale quantum computation.
DQC not only reduces circuit complexity and overall depth while mitigating inter-qubit crosstalk \cite{lai2025realization}, but also offers a scalable platform for the implementation of fault-tolerant quantum computation \cite{kim2024fault,akahoshi2024partially,aghaee2025scaling} and measurement-based quantum computing \cite{gross2007novel}.

Long-distance quantum gates, also known as quantum gate teleportation \cite{hu2023progress}, have been recognized as a critical technology for constructing DQC \cite{lim2005repeat,jiang2007distributed,oh2023distributed}, large-scale quantum network, teleportation-based quantum communication, and universal quantum computation \cite{gottesman1999demonstrating}.
Similar to quantum state teleportation, gate teleportation excutes nonlocal quantum gate across spatially separated network nodes, escaping the direct interactions between distant gate qubits \cite{eisert2000optimal}.
In many scenarios, gate teleportation often achieves higher efficiency than conventional quantum state teleportation \cite{lv2024demonstration}.
Long-distance two-qubit controlled-NOT (CNOT) gates were first experimentally demonstrated in photonic systems \cite{huang2004experimental,gao2010teleportation}, and were later experimentally realized in superconducting circuits \cite{chou2018deterministic}, trapped-ion systems \cite{wan2019quantum}, and atom-cavity systems \cite{daiss2021quantum}.
Recently, many long-distance CNOT gates have been reported across a variety of physical platforms, including a photonic CNOT gate over 7 $km$ of optical fiber \cite{liu2024nonlocal}, a 1 $km$ chip-to-chip photonic CNOT gate \cite{Feng2024}, a 2 $m$ trapped-ion CNOT gate \cite{main2024distributed}, and a 64 $m$ superconducting CNOT gate \cite{qiu2025deterministic}.
In addition, a photonic CNOT gate utilizing indefinite causal order has been  proposed \cite{liu2025quantum,liu2026NJP}.
Despite these advances, long-distance quantum gates involving more qubits have received limited theoretical and experimental attention to date.

Implementing multi-qubit gates is one of the essential problems in large-scale quantum computing.
Besides the multi-qubit Toffoli gate \cite{lanyon2009simplifying}, which involves multiple control qubits acting on a single target qubit, the multi-qubit fan-out gate also plays a particularly important role. The fan-out gate conditionally flips all target qubits when the one common control qubit is in the state $|1\rangle$, while leaving them unchanged otherwise.
Fan-out gates have been widely used in quantum computing \cite{gokhale2021quantum}, quantum error-correction circuit \cite{postler2022demonstration}, quantum state generation \cite{su2022single,zhao2023generation,li2024heralded,liu2025deterministic}, and the construction of classical functions.
Nowadays, local fan-out gates have been experimentally realized in Rydberg atom systems \cite{khazali2020fast,young2021asymmetric,wei2022fast}, photonic platforms \cite{xu2023multiqubit}, and superconducting systems \cite{song2025constant,hashim2025efficient}.
It is well-known that a general $n$-qubit fan-out gate, denoted CNOT$^{n-1}$ or $C^{n}$, can be synthesized by using $(n-1)$ CNOT gates. The nonlocal $C^n$ gate can be constructed by sharing multiple Greenberger-Horne-Zeilinger (GHZ) or Bell states using the decomposition-based approach \cite{yimsiriwattana2004generalized}.
However, such inefficient synthesis makes the gates more vulnerable to environmental noise and imposes substantial resource overhead in practical scenarios.
This motivates the search for more efficient approaches to implement long-distance fan-out gates beyond the standard gate synthesis. Fortunately, independent of the gate synthesis strategy, a long-distance four-qubit fan-out gate is constructed recently using a GHZ state and reduces circuit depth for practical applications \cite{fanout2026}.
Notably,  most pioneering demonstrations of long-distance gates \cite{huang2004experimental,liu2024nonlocal,Feng2024,xu2023multiqubit} have relied on destructive measurements of certain output photons, which limits scalability and hinders integration into larger quantum architectures.
 Although heralded and nondestructive CNOT gates have been reported using entangled photon pairs \cite{pittman2001probabilistic,zhao2005experimental,zeuner2018integrated} or single photons \cite{bao2007optical,li2021heralded,liu2023linear},
 nondestructive long-distance fan-out gates remain largely unexplored.

In this paper, we propose a simplified protocol for implementing a nondestructive long-distance three-qubit $C^3$ gate among three spatially separated parties, assisted by a pre-shared three-qubit GHZ state.
The protocol is then naturally extended to a general $n$-qubit fan-out gate by employing an $n$-qubit GHZ state and $n$ local CNOT operations.
Our scheme reduces the number of CNOT gates by $(n-2)$ compared with the standard  decomposition-based approach.
Subsequently, we propose two optical architectures for implementing heralded and post-selected $C^n$ gates, respectively. In the heralded scheme, the success is confirmed by coincidence measurements on the ancillary photons, while in the post-selected scheme, the success is identified by events in which exactly one photon occupies each output spatial mode. The total success probabilities of the heralded and post-selected protocols are $(\frac{1}{2})^{3n-1}$ and $(\frac{1}{2})^{\frac{5n}{2}}$ for even $n$ or $(\frac{1}{2})^{\frac{5n-3}{2}}$ for odd $n$, respectively, which beat the decomposition-based approach.
Finally, we evaluate the performance of our $n$-qubit fan-out gate and show that it maintains high fidelity under low depolarizing noise.
These results show the potential of our scheme as a scalable building block for distributed quantum computing.

\section{Nondestructive long-distance $n$-qubit fan-out gate}  \label{Sec2}

\begin{figure}   [!h]
\begin{center}
\includegraphics[width=8.5 cm,angle=0]{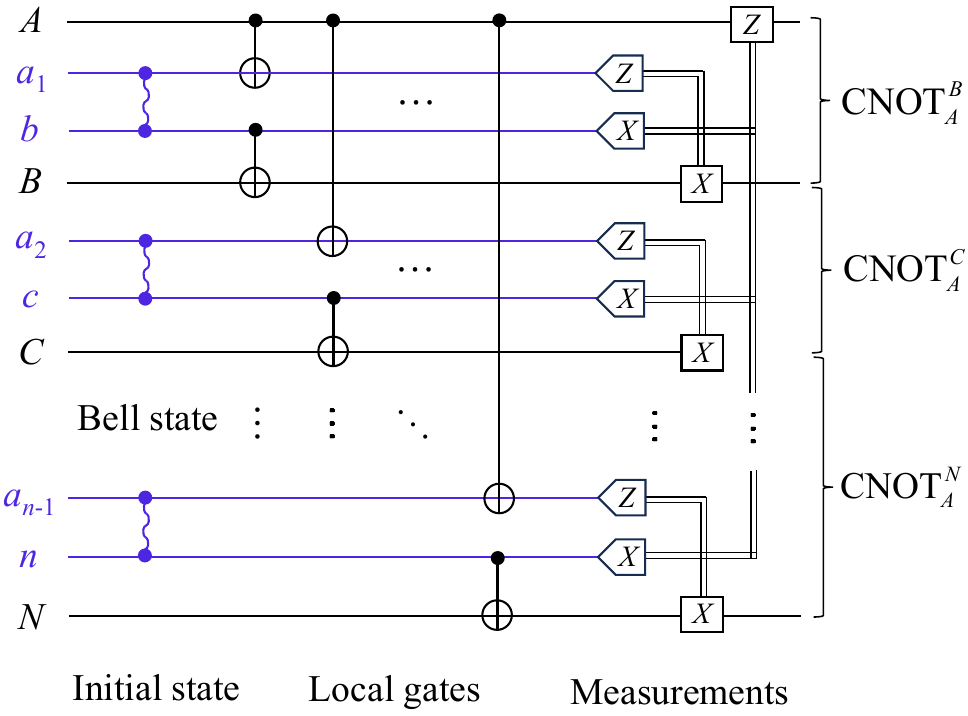}
\caption{
\justifying
Quantum circuit for implementing an $n$-qubit long-distance $C^n$ gate acting on $n$ remote qubits $A, B, \ldots, N$, based on the standard decomposition-based method \cite{baumer2025measurement}. Here $\mathrm{CNOT}_i^j$ denotes a CNOT gate with control qubit $i$ and target qubit $j$. Qubits $A, a_1, \ldots, a_{n-1}$ are held by Alice, qubits $B$ and $b$ by Bob, qubits $C$ and $c$ by Charlie, and so on.}
\label{decomposion}
\end{center}
\end{figure}

A widely adopted strategy for implementing a nondestructive long-distance $n$-qubit fan-out gate is to decompose the $C^n$ operation into $(n-1)$ long-distance two-qubit CNOT gates \cite{baumer2025measurement}.
In this standard decomposition-based method, each long-distance CNOT gate can be constructed using one pre-shared Bell state  and two local CNOT gates \cite{eisert2000optimal}.
Figure~\ref{decomposion} shows a standard decomposition-based long-distance $C^n$ gate with one control qubit $A$ and $(n-1)$ target qubits $B, C, \ldots, N$ can be synthesized using $(n-1)$ Bell states and $2(n-1)$ local CNOT gates \cite{baumer2025measurement}.

However, the decomposition-based method requires a relatively large amount of entanglement resources and CNOT gates.
We next present our simpler schemes for constructing long-distance $C^n$ gates. We begin with the three-qubit $C^3$ gate as a basic example.

\subsection{Long-distance three-qubit fan-out gate}  \label{Sec2.1}

\begin{figure} %[!h]   %[htpb]
\begin{center}
\includegraphics[width=8 cm,angle=0]{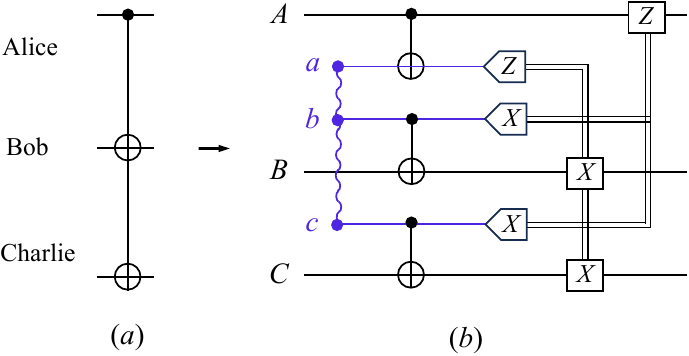}
\caption{\justifying (a) Circuit representation of the three-qubit fan-out $C^3$ gate.
(b) Quantum circuit for implementing a long-distance $C^3$ gate on three spatially separated qubits $A$, $B$ and $C$ held by Alice, Bob, and Charlie, respectively.
A pre-shared  ancillary GHZ state acts as the quantum channel and interacts with the gate qubits via three local CNOT gates.
The nonlocal $C^3$  gate is completed by performing single-qubit measurements on the ancillary qubits in appropriate bases, followed by classical feed-forward operations.
} \label{nonlocalC3}
\end{center}
\end{figure}

Three-qubit fan-out gate $C^3$ is the simplest nontrivial fan-out quantum gate, and it can be written as
\begin{eqnarray}           \label{eq1}
U_{C^3}=|0\rangle_c\langle0|\otimes I_{t_1}\otimes I_{t_2}+ |1\rangle_c\langle1|\otimes X_{t_1}\otimes X_{t_2}.
\end{eqnarray}
Here $I_{t_i}$ and $X_{t_i}$ denote the $2 \times 2$ identity and Pauli-$X$ operators acting on the $i$th target qubit ($i = 1, 2$), respectively. As illustrated in figure~\ref{nonlocalC3}(a), this gate functions as a controlled-NOT-NOT operation. If the control qubit is in the state $|1\rangle$, both target qubits undergo a qubit-flip operation, while no action is applied if the control qubit is in the state $|0\rangle$.

Consider there are three parties, Alice, Bob, and Charlie, located at spatially separated sites.
They aim to implement a long-distance $C^3$ gate, in which Alice serves as the control party, while Bob and Charlie serve as the target parties. To achieve this, as shown in figure \ref{nonlocalC3}(b), our scheme completes the three-qubit fan-out gate in just three steps.

In the first step, Alice, Bob, and Charlie share a pre-distributed three-qubit GHZ state $|\phi\rangle_{abc}$ as quantum channel
\begin{eqnarray}
\begin{split}             \label{eq2}
|\phi\rangle_{abc}=\frac{1}{\sqrt{2}}(|0\rangle_a|0\rangle_b|0\rangle_c+|1\rangle_a|1\rangle_b|1\rangle_c).
\end{split}
\end{eqnarray}
The above ancillary qubits $a$, $b$, and $c$ are possessed by Alice, Bob, and Charlie, respectively. Each party also holds a computational qubit $A$, $B$, and $C$ for Alice, Bob, and Charlie, respectively, and they are initially prepared in the following general pure states
\begin{eqnarray}
\begin{split}             \label{eq3}
|\phi\rangle_A =\alpha_1|0\rangle_A+\beta_1|1\rangle_A,
\end{split}
\end{eqnarray}
\begin{eqnarray}
\begin{split}             \label{eq4}
|\phi\rangle_{B} =\alpha_2|0\rangle_{B}+\beta_2|1\rangle_{B},
\end{split}
\end{eqnarray}
\begin{eqnarray}
\begin{split}             \label{eq5}
|\phi\rangle_{C} =\alpha_3|0\rangle_{C}+\beta_3|1\rangle_{C}.
\end{split}
\end{eqnarray}
Here the complex coefficients satisfy the normalization condition $|\alpha_i|^2+|\beta_i|^2=1$ ($i=1,2,3$).
For simplicity, we denote the state $|\phi\rangle_{ABC}=|\phi\rangle_{A}\otimes|\phi\rangle_{B}\otimes|\phi\rangle_{C}$.

In the second step, Alice  applies a local CNOT gate on qubits $A$ and $a$,  and Bob (Charlie) applies a local CNOT gate on qubits $B$ and $b$ ($C$ and $c$). The qubits $A$, $b$, $c$ act as the  control qubits, and qubits $a$, $B$, $C$ act as the target qubits. Above three CNOT gates transform  the whole system from initial state $|\Phi_0\rangle $ into $|\Phi_1\rangle $. Here
\begin{eqnarray}
\begin{split}             \label{eq6}
|\Phi_0\rangle =|\phi\rangle_{abc}\otimes|\phi\rangle_A\otimes|\phi\rangle_B\otimes|\phi\rangle_C,
\end{split}
\end{eqnarray}
\begin{eqnarray}
\begin{split}             \label{eq7}
|\Phi_1\rangle =&\;\frac{1}{2\sqrt{2}}
 \big(I_AI_BI_CC^3_{ABC}|\phi\rangle_{ABC}\otimes|0\rangle_a|+\rangle_b|+\rangle_c
+Z_AI_BI_CC^3_{ABC}|\phi\rangle_{ABC}\otimes|0\rangle_a|+\rangle_b|-\rangle_c
\\&\;+Z_AI_BI_CC^3_{ABC}|\phi\rangle_{ABC}\otimes|0\rangle_a|-\rangle_b|+\rangle_c
+I_AI_BI_CC^3_{ABC}|\phi\rangle_{ABC}\otimes|0\rangle_a|-\rangle_b|-\rangle_c
\\&\;+I_AX_BX_CC^3_{ABC}|\phi\rangle_{ABC}\otimes|1\rangle_a|+\rangle_b|+\rangle_c
-Z_AX_BX_CC^3_{ABC}|\phi\rangle_{ABC}\otimes|1\rangle_a|+\rangle_b|-\rangle_c
\\&\;-Z_AX_BX_CC^3_{ABC}|\phi\rangle_{ABC}\otimes|1\rangle_a|-\rangle_b|+\rangle_c
+I_AX_BX_CC^3_{ABC}|\phi\rangle_{ABC}\otimes|1\rangle_a|-\rangle_b|-\rangle_c\big).
\end{split}
\end{eqnarray}
Here the $I_j=|0\rangle_j\langle0|+|1\rangle_j\langle1|$, $Z_j=|0\rangle_j\langle1|-|1\rangle_j\langle0|$, and $X_j=|0\rangle_j\langle1|+|1\rangle_j\langle0|$ are identity, Pauli $Z$, and Pauli $X$ operations on qubit $j$ ($j=A,B,C$).

In the third step, single-qubit measurements are performed on qubit $a$ in the basis $\{|0\rangle, |1\rangle\}$, and qubits $b$, and $c$ are performed in the basis $\{|\pm\rangle\}$, where $|\pm\rangle = \frac{1}{\sqrt{2}}(|0\rangle \pm |1\rangle)$.

Finally, as indicated by equation~\eqref{eq7}, after performing the three single-qubit projective measurements and exchanging the measured results via classical communication, appropriate feed-forward operations are applied to qubits $A$, $B$, and $C$ to complete the $C^3$ gate. More specifically, the correspondences between the measurement outcomes and the required local Pauli corrections are summarized in table \ref{Table1}.

\begin{table} [tb] \renewcommand{\arraystretch}{1.1}  % 增大行间距
\centering \caption{\justifying The correspondence between measured outcomes on the qubits $a$, $b$, and $c$  and the classical feed-forward operations on qubits $A$, $B$, and $C$  to complete the three-qubit $C^3$ gate. The symbol “\textbf{--}” indicates that no correction operation is required. $Z=|0\rangle\langle1|-|1\rangle\langle0|$ and $X=|0\rangle\langle1|+|1\rangle\langle0|$ are the Pauli $Z$ and Pauli $X$ operations.}
\begin{tabular}{cccccccccccccc}%{lp{1.2in} lp{1.2in}}
\hline

\multicolumn {3}{c}{Measurement}  &     & \multicolumn {3}{c}{\;  Feed-forward}     \\ \cline{1-3}  \cline{5-7}

    Qubit $a$   &  \,    Qubit $b$    &  \,    Qubit $c$      &      & \,  Qubit $A$   & \,   Qubit $B$     &  \, Qubit $C$                   \\

\hline

$|0\rangle_{a}$  & \; $|+\rangle_{b}$  & \; $|+\rangle_{c}$   &      & \;  --    &  \;  --     &   \; --                    \\

$|0\rangle_{a}$  & \; $|+\rangle_{b}$  & \; $|-\rangle_{c}$   &      & \;  $Z$    &  \; --     &  \;  --                    \\

$|0\rangle_{a}$  & \; $|-\rangle_{b}$  & \; $|+\rangle_{c}$   &    & \; $Z$    &  \;  --     &  \; --                    \\

$|0\rangle_{a}$  & \; $|-\rangle_{b}$  & \; $|-\rangle_{c}$   &      & \;  --   &  \; --    &   \; --                    \\

$|1\rangle_{a}$  & \; $|+\rangle_{b}$  & \; $|+\rangle_{c}$   &      & \; --   &  \;   $X$      &  \; $X$                   \\

$|1\rangle_{a}$  & \; $|+\rangle_{b}$  & \; $|-\rangle_{c}$   &      & \; $Z$      &  \;  $X$       &  \; $X$                   \\

$|1\rangle_{a}$  & \; $|-\rangle_{b}$  & \; $|+\rangle_{c}$   &      & \; $Z$      &  \; $X$      &  \; $X$                    \\

$|1\rangle_{a}$  & \; $|-\rangle_{b}$  & \; $|-\rangle_{c}$   &      & \; --      &  \;  $X$       &  \; $X$                     \\

                             \hline
\end{tabular}\label{Table1}
\end{table}

\subsection{Long-distance $n$-qubit fan-out gate}  \label{Sec2.2}

\begin{figure}   [tpb]%[!h]
\begin{center}
\includegraphics[width=8 cm,angle=0]{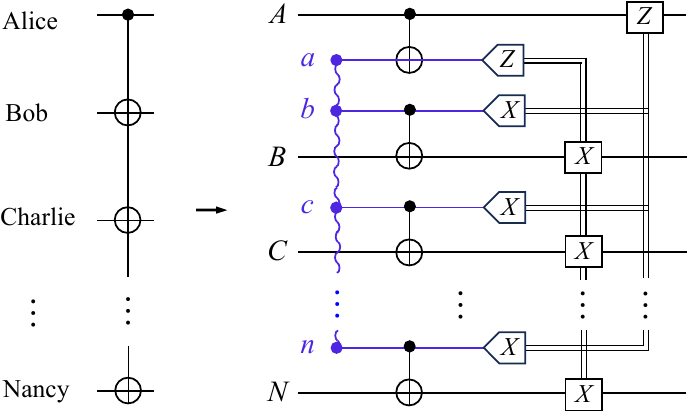}
\caption{\justifying Quantum circuit for implementing a long-distance $C^n$ gate on $n$ remote qubits $A$, $B$,  $\dots, N$.
An $n$-qubit GHZ state as ancillary quantum channel and $n$ local CNOT gates are required.}  \label{nonlocalCN}
\end{center}
\end{figure}

Our scheme shown in figure~\ref{nonlocalC3} can be naturally extended to implement a long-distance general $n$-qubit fan-out  $C^n$ gate.
As illustrated in figure~\ref{nonlocalCN}, we consider an initial system composed of $n$ spatially separated computational qubits $A$, $B$, $\dots$, $N$, which are prepared in an arbitrary initial state
\begin{eqnarray}
\begin{split}             \label{eq8}
|\phi\rangle_{AB\dots N} = &\;c_1|0\rangle^{\otimes n}+c_2|0\rangle^{\otimes (n-1)}|1\rangle
+ \dots  +c_{2^n}|1\rangle^{\otimes n}.
\end{split}
\end{eqnarray}
Here $c_l$ ($l=1, \ldots, 2^n$) are normalized complex coefficients. Qubit $A$ acts as the control qubit, while qubits $B$, $C$, $\dots$, $N$ act as the target qubits. The qubits $A$, $B$, $\dots$, $N$ are held by $n$ spatially separated parties, Alice, Bob, $\dots$, and Nancy, respectively.
Let us now describe the protocol for implementing the $C^n$ gate step by step.

Firstly, to implement the long-distance $C^n$ gate, an $n$-qubit pre-shared GHZ state among $n$ parties as the quantum channel is required. This GHZ state can be generated from the initial product state $|0\rangle_a \otimes |0\rangle_b \cdots \otimes |0\rangle_n$ by applying one Hadamard gate and $(n-1)$ CNOT gates. The ancillary $n$-qubit GHZ state is explicitly given by
\begin{eqnarray}
\begin{split}             \label{eq9}
|\phi\rangle_{ab\dots n}=\frac{1}{\sqrt{2}}(|0\rangle^{\otimes n}+|1\rangle^{\otimes n})_{ab\dots n}.
\end{split}
\end{eqnarray}
The qubits $a$, $b$, $\dots$, $n$ are held by  Alcie, Bob, $\dots$, Nancy, respectively.

Secondly, as shown in figure \ref{nonlocalCN}, Alice, Bob, $\ldots$, and Nancy apply local CNOT gates to their individual qubit pairs  $(A,a)$, $(B,b)$, $\ldots$, $(N,n)$,  respectively.
These CNOT operations transform the state of the system from $|\Psi_0\rangle$ to $|\Psi_1\rangle$. $|\Psi_0\rangle$ and $|\Psi_1\rangle$ are given by
\begin{eqnarray}
\begin{split}             \label{eq10}
|\Psi_0\rangle =|\phi\rangle_{AB\dots N}\otimes|\phi\rangle_{ab\dots n},
\end{split}
\end{eqnarray}
\begin{eqnarray}
\begin{split}             \label{eq11}
|\Psi_1\rangle =&\;\frac{1}{\sqrt{2^n}}
\big[
   I_A I_B \dots I_N C^N_{A\dots N}   |\phi\rangle_{AB\dots N} \otimes |0\rangle_a \sum_{\vec{\mu} \in \mathcal{E}} |\vec{\mu}\rangle \\
&+ Z_A I_B \dots I_N C^N_{AB\cdots N} |\phi\rangle_{AB\dots N} \otimes |0\rangle_a \sum_{\vec{\mu} \in \mathcal{O}} |\vec{\mu}\rangle \\
&+ I_A X_B \dots X_N C^N_{AB\cdots N} |\phi\rangle_{AB\dots N} \otimes |1\rangle_a \sum_{\vec{\mu} \in \mathcal{E}} |\vec{\mu}\rangle \\
&- Z_A X_B \dots X_N C^N_{AB\cdots N} |\phi\rangle_{AB\dots N} \otimes |1\rangle_a \sum_{\vec{\mu} \in \mathcal{O}} |\vec{\mu}\rangle
\big].
\end{split}
\end{eqnarray}
Here $\vec{\mu} = (\mu_b, \mu_c, \ldots, \mu_n)$ is ancillary qubits $b, \ldots, n$ with each $\mu_j \in \{+, -\}$,
and  $|\vec{\mu}\rangle = |\mu_b\rangle |\mu_c\rangle \ldots |\mu_n\rangle$.
The summation indices $\vec{\mu} \in \mathcal{E}$ and $\vec{\mu} \in \mathcal{O}$ run over all bitstrings containing an even and odd number of $|-\rangle$ states, respectively.

Thirdly, the ancillary qubit $a$ held by Alice is measured in the basis $\{|0\rangle, |1\rangle\}$, while the remaining ancillary qubits $b, \dots, n$ are measured in the basis $\{|\pm\rangle\}$, respectively.

Finally, based on equation~\eqref{eq11}, after performing single-qubit projective measurements on the ancillary qubits, the parties exchange their measurement outcomes via classical communication. The detector-dependent outcomes determine the feed-forward operations required to complete the $C^n$ gate. Specifically,

(i)  If the measurement outcome is $|0\rangle_a$ and the number of $|-\rangle$ outcomes among ${b, \ldots, n}$ is even, then no correction is needed.

(ii)  If the measurement outcome is  $|0\rangle_a$ and the number of $|-\rangle$ outcomes among ${b, \ldots, n}$ is odd, then Alice applies a $Z$ operation on qubit $A$.

(iii) If the measurement outcome is $|1\rangle_a$ and the number of $|-\rangle$ outcomes among ${b, \ldots, n}$ is even, then Bob to Nancy apply $X$ operations on qubits $B$  to $N$, respectively.

(iv) If the measurement outcome is $|1\rangle_a$ and the number of $|-\rangle$ outcomes among ${b, \ldots, n}$ is odd, then Alice applies a $Z$ operation on qubit $A$, and Bob to Nancy apply $X$ operations on qubits $B$ to $N$, respectively.

Our scheme, shown in figure~\ref{nonlocalCN}, requires one $n$-qubit GHZ state and $n$ local CNOT gates to implement a long-distance $C^n$ gate. The $C^n$ gate is nondestructive because its success is  heralded by measuring only the ancillary qubits, and the computational qubits are kept for subsequent applications. Furthermore, the $n$-qubit GHZ state can be prepared using $(n-1)$ CNOT gates and one Hadamard gate. Therefore, our approach for implementing a long-distance $C^n$ gate requires a total of $(2n-1)$ local CNOT gates and one Hadamard gate.
In contrast, constructing a long-distance $C^n$ gate via the standard decomposition-based method (Fig.~\ref{decomposion}) requires $3(n-1)$ local CNOT gates and $(n-1)$ Hadamard gates \cite{baumer2025measurement}, since each Bell state can be created using one CNOT gate and one Hadamard gate. A detailed comparison of gate counts between our scheme and previous approaches is presented in table~\ref{Table3}.

\begin{table} [!h] \renewcommand{\arraystretch}{1.1}  % 增大行间距
\centering \caption{\justifying Comparisons between our scheme and previous counterparts for implementing a long-distance $C^n$ gate in terms of circuit gate count.}
\begin{tabular}{cccccccccccccc}%{lp{1.2in} lp{1.2in}}

\hline

    Schemes                        &&\quad    CNOT   gate         &&\quad  Hadamard  gate    \\

\hline

Decomposition-based method with Ref. \cite{baumer2025measurement}  &&\quad        $3n-3$          &&\quad $n-1$             \\

The method with Ref. \cite{song2025constant}       &&\quad         $3n-3$         &&\quad  $2n-2$             \\

Ours                               &&\quad        $2n-1$          &&\quad   $1$               \\

\hline
\end{tabular}\label{Table3}
\end{table}

\section{Implementations of long-distance fan-out gate with linear optics}  \label{Sec3}

The linear-optical fan-out gate plays an important role in optical quantum computing, yet its realizations remain largely unexplored. In this section, we will explore an approach to implement such gates with an improved success probability.

\subsection{Heralded long-distance $C^3$ gate}  \label{Sec3.1}

In this subsection, based on the approaches presented in section \ref{Sec2.1}, we propose a heralded and nondestructive protocol for a long-distance $C^3$ gate using linear optics. Here ``heralded" means that the success of the scheme can be confirmed by solely measuring the ancillary modes.

\begin{figure}   [tpb]%[!h]
\begin{center}
\includegraphics[width=8.3 cm,angle=0]{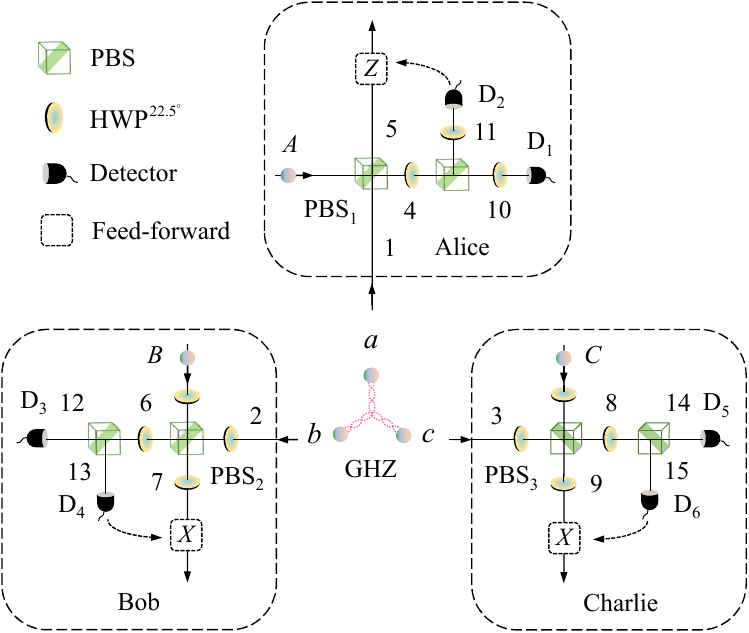}
\caption{\justifying Schematic setup for implementing the heralded and nondestructive long-distance $C^3$ gate acting on the polarization states of photons in spatial modes 5, 7, and 9.
PBS denotes a polarizing beam splitter that transmits $H$-polarized photons and reflects $V$-polarized photons.
HWP$^{22.5^\circ}$ represents a half-wave plate set at $22.5^\circ$, performing the transformations $|H\rangle \rightarrow (|H\rangle + |V\rangle)/\sqrt{2}$ and $|V\rangle \rightarrow (|H\rangle - |V\rangle)/\sqrt{2}$.
Numbers label photon spatial modes, and D$_i$ denotes a single-photon detector.}
\label{optical3}
\end{center}
\end{figure}

As illustrated in figure \ref{optical3}, the qubits are encoded in the polarization states of single photons, i.e., $|H\rangle \equiv| 0\rangle$,  $|V\rangle \equiv| 1\rangle$. Here $|H\rangle$ and $|V\rangle$ denote  horizontal and  vertical polarization components, respectively.
The pre-shared GHZ state given in equation \eqref{eq2} is expressed in polarization-encoded form as
\begin{eqnarray}
\begin{split}             \label{eq12}
|\phi'\rangle_{abc}=\frac{1}{\sqrt{2}}(|H\rangle_a|H\rangle_b|H\rangle_c+|V\rangle_a|V\rangle_b|V\rangle_c).
\end{split}
\end{eqnarray}
The initial states of the three computational qubits $A$, $B$, and $C$ given in equations \eqref{eq3}-\eqref{eq5} correspond to
\begin{eqnarray}
\begin{split}             \label{eq13}
|\phi'\rangle_A =\alpha_1|H\rangle_A+\beta_1|V\rangle_A,
\end{split}
\end{eqnarray}
\begin{eqnarray}
\begin{split}             \label{eq14}
|\phi'\rangle_{B} =\alpha_2|H\rangle_{B}+\beta_2|V\rangle_{B},
\end{split}
\end{eqnarray}
\begin{eqnarray}
\begin{split}             \label{eq15}
|\phi'\rangle_{C} =\alpha_3|H\rangle_{C}+\beta_3|V\rangle_{C}.
\end{split}
\end{eqnarray}
Thus, the total input state of the system corresponding to equation~\eqref{eq6}  is
\begin{eqnarray}
\begin{split}             \label{eq16}
|\Phi'_0\rangle =|\phi'\rangle_{abc}\otimes|\phi'\rangle_{A}\otimes|\phi'\rangle_{B}\otimes|\phi'\rangle_{C}.
\end{split}
\end{eqnarray}

Firstly, at Alice's station, photons $A$ and $a$ interfere at a polarizing beam splitter (PBS$_1$), which transmits $H$-polarized photons and reflects $V$-polarized photons. Specifically, the $H$-polarized photon emitted from mode $A$ leads to mode 4 and the $V$-polarized photon to mode 5, while the $H$-polarized photon from mode $a$ leads to mode 5 and the $V$-polarized photon to mode 4. Consequently, after PBS$_1$, the initial state $|\Phi'_0\rangle$ is converted into
\begin{eqnarray}
\begin{split}             \label{eq17}
|\Phi'_1\rangle=&\;\frac{1}{\sqrt{2}}(\alpha_1 |H\rangle_4 |H\rangle_5 |H\rangle_2 |H\rangle_3
                   +\alpha_1 |H\rangle_4 |V\rangle_4 |V\rangle_2 |V\rangle_3
                   + \beta_1 |V\rangle_5 |H\rangle_5 |H\rangle_2 |H\rangle_3\\&\;
                   + \beta_1 |V\rangle_5 |V\rangle_4 |V\rangle_2 |V\rangle_3)
             \otimes(\alpha_2|H\rangle_{B} + \beta_2|V\rangle_{B})
             \otimes(\alpha_3|H\rangle_{C} + \beta_3|V\rangle_{C}).
\end{split}
\end{eqnarray}
Here the $|H\rangle_i$ ($|V\rangle_i$) denotes that the $H$-polarized ($V$-polarized) photon is in the spatial mode $i$.

In the equation \eqref{eq17}, the items
$|H\rangle_4 |H\rangle_5 |H\rangle_2 |H\rangle_3$ and
$|V\rangle_5 |V\rangle_4 |V\rangle_2 |V\rangle_3$ represent configurations in which each of the spatial modes 4 and 5 contains exactly one photon.
In contrast, the item $|H\rangle_4 |V\rangle_4 |V\rangle_2 |V\rangle_3$ indicates that two photons occupy spatial mode 4, and the item $|V\rangle_5 |H\rangle_5 |H\rangle_2 |H\rangle_3$ indicates that two photons occupy spatial mode 5.
By postselecting events in which modes 4 and 5 each contain exactly one photon, the state $|\Phi'_1\rangle$ collapses to
\begin{eqnarray}
\begin{split}             \label{eq18}
|\Phi'_2\rangle=&\;(\alpha_1|H\rangle_2|H\rangle_3|H\rangle_4|H\rangle_5
                     +\beta_1|V\rangle_2|V\rangle_3|V\rangle_4|V\rangle_5)
                     \otimes(\alpha_2|H\rangle_{B}+\beta_2|V\rangle_{B}) \\&\;
                      \otimes(\alpha_3|H\rangle_{C}+\beta_3|V\rangle_{C}),
\end{split}
\end{eqnarray}
with a probability of $\frac{1}{2}$.

Secondly, at Bob's and Charlie's stations, the photons $B$ and $b$ emitted from spatial mode 2, and the photons $C$ and $c$ emitted from spatial mode 3, go through half-wave plates set at $22.5^\circ$ (HWPs$^{22.5^\circ}$). The HWP$^{22.5^\circ}$ implements the Hadamard operation on the polarization degree of freedom (DOF), i.e.,
\begin{eqnarray}
\begin{split}             \label{eq18.1}
|H\rangle \xrightarrow{\text{HWP}^{22.5^\circ}} |D\rangle =\frac{1}{\sqrt{2}}(|H\rangle + |V\rangle),  \quad
|V\rangle \xrightarrow{\text{HWP}^{22.5^\circ}} |A\rangle =\frac{1}{\sqrt{2}}(|H\rangle - |V\rangle).
\end{split}
\end{eqnarray}
The four HWPs$^{22.5^\circ}$ in spatial modes $B$, 2, $C$, and 3 transform the state $|\Phi'_2\rangle$ into
\begin{eqnarray}
\begin{split}             \label{eq18.2}
|\Phi'_3\rangle=&\;\frac{1}{4}[\alpha_1(|H\rangle_2+|V\rangle_2)(|H\rangle_3+|V\rangle_3)|H\rangle_4|H\rangle_5
                     +\beta_1(|H\rangle_2-|V\rangle_2)(|H\rangle_3-|V\rangle_3)|V\rangle_4|V\rangle_5]\\&\;
                     \otimes[\alpha_2(|H\rangle_{B}+|V\rangle_{B})+\beta_2(|H\rangle_{B}-|V\rangle_{B})]
                      \otimes[\alpha_3(|H\rangle_{C}+|V\rangle_{C})+\beta_3(|H\rangle_{C}-|V\rangle_{C})].
\end{split}
\end{eqnarray}

Thirdly, the photons emitted from the spatial modes $B$ and 2 simultaneously arrive at PBS$_2$, while the photons emitted from spatial modes $C$ and 3 simultaneously arrive at PBS$_3$. The two PBS$_2$ and PBS$_2$ change state $|\Phi'_3\rangle$ into
\begin{eqnarray}
\begin{split}             \label{eq18.3}
|\Phi'_4\rangle=&\;\frac{1}{4}[\alpha_1(|H\rangle_6+|V\rangle_7)(|H\rangle_8+|V\rangle_9)|H\rangle_4|H\rangle_5
                     +\beta_1(|H\rangle_6-|V\rangle_7)(|H\rangle_8-|V\rangle_9)|V\rangle_4|V\rangle_5]\\&\;
                     \otimes[\alpha_2(|H\rangle_7+|V\rangle_6)+\beta_2(|H\rangle_7-|V\rangle_6)]
                      \otimes[\alpha_3(|H\rangle_9+|V\rangle_8)+\beta_3(|H\rangle_9-|V\rangle_8)].
\end{split}
\end{eqnarray}
To clearly illustrate the optical scheme, equation~\eqref{eq18.3} is expanded explicitly as
\begin{eqnarray}
\begin{split}             \label{eq18.4}
|\Phi'_4\rangle=&\;\frac{1}{4}[\alpha_1\alpha_2\alpha_3|H\rangle_4|H\rangle_5(|H\rangle_6|H\rangle_8+|V\rangle_7|H\rangle_8
                 +|H\rangle_6|V\rangle_9+|V\rangle_7|V\rangle_9)(|H\rangle_7|H\rangle_9 \\&\;
                 +|V\rangle_6|H\rangle_9+|H\rangle_7|V\rangle_8+|V\rangle_6|V\rangle_8)
                 %%%%%
                 +\alpha_1\alpha_2\beta_3|H\rangle_4|H\rangle_5(|H\rangle_6|H\rangle_8+|V\rangle_7|H\rangle_8\\&\;
                 +|H\rangle_6|V\rangle_9+|V\rangle_7|V\rangle_9)(|H\rangle_7|H\rangle_9
                 +|V\rangle_6|H\rangle_9-|H\rangle_7|V\rangle_8-|V\rangle_6|V\rangle_8)\\&\;
                 %%%%%
                 +\alpha_1\beta_2\alpha_3|H\rangle_4|H\rangle_5(|H\rangle_6|H\rangle_8+|V\rangle_7|H\rangle_8
                 +|H\rangle_6|V\rangle_9+|V\rangle_7|V\rangle_9)(|H\rangle_7|H\rangle_9\\&\;
                 -|V\rangle_6|H\rangle_9+|H\rangle_7|V\rangle_8-|V\rangle_6|V\rangle_8)
                 %%%%%
                 +\alpha_1\beta_2\beta_3|H\rangle_4|H\rangle_5(|H\rangle_6|H\rangle_8+|V\rangle_7|H\rangle_8\\&\;
                 +|H\rangle_6|V\rangle_9+|V\rangle_7|V\rangle_9)(|H\rangle_7|H\rangle_9
                 -|V\rangle_6|H\rangle_9-|H\rangle_7|V\rangle_8+|V\rangle_6|V\rangle_8)\\&\;
                 %%%%%
                 +\beta_1\alpha_2\alpha_3|V\rangle_4|V\rangle_5(|H\rangle_6|H\rangle_8-|V\rangle_7|H\rangle_8
                 -|H\rangle_6|V\rangle_9+|V\rangle_7|V\rangle_9)(|H\rangle_7|H\rangle_9\\&\;
                 +|V\rangle_6|H\rangle_9+|H\rangle_7|V\rangle_8+|V\rangle_6|V\rangle_8)
                 %%%%%
                 +\beta_1\alpha_2\beta_3|V\rangle_4|V\rangle_5(|H\rangle_6|H\rangle_8-|V\rangle_7|H\rangle_8\\&\;
                 -|H\rangle_6|V\rangle_9+|V\rangle_7|V\rangle_9)(|H\rangle_7|H\rangle_9
                 +|V\rangle_6|H\rangle_9-|H\rangle_7|V\rangle_8-|V\rangle_6|V\rangle_8)\\&\;
                 %%%%%
                 +\beta_1\beta_2\alpha_3|V\rangle_4|V\rangle_5(|H\rangle_6|H\rangle_8-|V\rangle_7|H\rangle_8
                 -|H\rangle_6|V\rangle_9+|V\rangle_7|V\rangle_9)(|H\rangle_7|H\rangle_9 \\&\;
                 -|V\rangle_6|H\rangle_9+|H\rangle_7|V\rangle_8-|V\rangle_6|V\rangle_8)
                 %%%%%
                 +\beta_1\beta_2\beta_3|V\rangle_4|V\rangle_5(|H\rangle_6|H\rangle_8-|V\rangle_7|H\rangle_8\\&\;
                 -|H\rangle_6|V\rangle_9+|V\rangle_7|V\rangle_9)(|H\rangle_7|H\rangle_9
                 -|V\rangle_6|H\rangle_9-|H\rangle_7|V\rangle_8+|V\rangle_6|V\rangle_8)].
\end{split}
\end{eqnarray}
Each term in equation~\eqref{eq18.4}  involves two, one, or no photons in spatial modes 6, 7, 8, and 9.
By postselecting the events in which each of these spatial modes contains exactly one photon, the state $|\Phi'_4\rangle$ in equation~\eqref{eq18.4} is transformed into
\begin{eqnarray}
\begin{split}             \label{eq18.5}
|\Phi'_5\rangle=&\;\frac{1}{2}[\alpha_1\alpha_2\alpha_3|H\rangle_4|H\rangle_5(|H\rangle_6|H\rangle_7 +|V\rangle_6|V\rangle_7)
                  (|H\rangle_8|H\rangle_9+|V\rangle_8|V\rangle_9)
                 %%%%%
                 +\alpha_1\alpha_2\beta_3|H\rangle_4|H\rangle_5\\&\;\otimes(|H\rangle_6|H\rangle_7+|V\rangle_6|V\rangle_7)
                  (|H\rangle_8|H\rangle_9-|V\rangle_8|V\rangle_9)
                 %%%%%
                 +\alpha_1\beta_2\alpha_3|H\rangle_4|H\rangle_5(|H\rangle_6|H\rangle_7\\&\;-|V\rangle_6|V\rangle_7)
                  (|H\rangle_8|H\rangle_9+|V\rangle_8|V\rangle_9)
                 %%%%%
                 +\alpha_1\beta_2\beta_3|H\rangle_4|H\rangle_5(|H\rangle_6|H\rangle_7 -|V\rangle_6|V\rangle_7)\\&\;
                  \otimes(|H\rangle_8|H\rangle_9-|V\rangle_8|V\rangle_9)
                 %%%%%
                 +\beta_1\alpha_2\alpha_3|V\rangle_4|V\rangle_5(|H\rangle_6|H\rangle_7 -|V\rangle_6|V\rangle_7)
                  (|H\rangle_8|H\rangle_9\\&\;-|V\rangle_8|V\rangle_9)
                 %%%%%
                 +\beta_1\alpha_2\beta_3|V\rangle_4|V\rangle_5(|H\rangle_6|H\rangle_7-|V\rangle_6|V\rangle_7)
                  (|H\rangle_8|H\rangle_9+|V\rangle_8|V\rangle_9)\\&\;
                 %%%%%
                 +\beta_1\beta_2\alpha_3|V\rangle_4|V\rangle_5(|H\rangle_6|H\rangle_7+|V\rangle_6|V\rangle_7)
                  (|H\rangle_8|H\rangle_9-|V\rangle_8|V\rangle_9)
                 %%%%%
                 +\beta_1\beta_2\beta_3|V\rangle_4|V\rangle_5\\&\;\otimes(|H\rangle_6|H\rangle_7+|V\rangle_6|V\rangle_7)
                  (|H\rangle_8|H\rangle_9+|V\rangle_8|V\rangle_9)],
\end{split}
\end{eqnarray}
with a probability of $\frac{1}{4}$.

Fourthly, the photons emitted from spatial modes 4, 6, 7, 8, and 9 pass through HWPs$^{22.5^\circ}$, respectively. These HWPs$^{22.5^\circ}$ transform the state $|\Phi'_5\rangle$ into
\begin{eqnarray}
\begin{split}             \label{eq18.6}
|\Phi'_6\rangle=&\;\frac{1}{2\sqrt{2}}[\alpha_1\alpha_2\alpha_3(|H\rangle_4+|V\rangle_4)|H\rangle_5(|H\rangle_6|H\rangle_7+|V\rangle_6|V\rangle_7)
                  (|H\rangle_8|H\rangle_9+|V\rangle_8|V\rangle_9)\\&\;
                 %%%%%
                 +\alpha_1\alpha_2\beta_3(|H\rangle_4+|V\rangle_4)|H\rangle_5(|H\rangle_6|H\rangle_7+|V\rangle_6|V\rangle_7)
                  (|H\rangle_8|V\rangle_9+|V\rangle_8|H\rangle_9)\\&\;
                 %%%%%
                 +\alpha_1\beta_2\alpha_3(|H\rangle_4+|V\rangle_4)|H\rangle_5(|H\rangle_6|V\rangle_7+|V\rangle_6|H\rangle_7)
                  (|H\rangle_8|H\rangle_9+|V\rangle_8|V\rangle_9)\\&\;
                 %%%%%
                 +\alpha_1\beta_2\beta_3(|H\rangle_4+|V\rangle_4)|H\rangle_5(|H\rangle_6|V\rangle_7+|V\rangle_6|H\rangle_7)
                  (|H\rangle_8|V\rangle_9+|V\rangle_8|H\rangle_9)\\&\;
                 %%%%%
                 +\beta_1\alpha_2\alpha_3(|H\rangle_4-|V\rangle_4)|V\rangle_5(|H\rangle_6|V\rangle_7+|V\rangle_6|H\rangle_7)
                  (|H\rangle_8|V\rangle_9+|V\rangle_8|H\rangle_9)\\&\;
                 %%%%%
                 +\beta_1\alpha_2\beta_3(|H\rangle_4-|V\rangle_4)|V\rangle_5(|H\rangle_6|V\rangle_7+|V\rangle_6|H\rangle_7)
                  (|H\rangle_8|H\rangle_9+|V\rangle_8|V\rangle_9)\\&\;
                 %%%%%
                 +\beta_1\beta_2\alpha_3(|H\rangle_4-|V\rangle_4)|V\rangle_5(|H\rangle_6|H\rangle_7+|V\rangle_6|V\rangle_7)
                  (|H\rangle_8|V\rangle_9+|V\rangle_8|H\rangle_9)\\&\;
                 %%%%%
                 +\beta_1\beta_2\beta_3(|H\rangle_4-|V\rangle_4)|V\rangle_5(|H\rangle_6|H\rangle_7+|V\rangle_6|V\rangle_7)
                  (|H\rangle_8|H\rangle_9+|V\rangle_8|V\rangle_9)].
\end{split}
\end{eqnarray}

Fifthly, the $H$-polarized photon from spatial mode 4 is transmitted to spatial mode 10 via a PBS, while the $V$-polarized photon is reflected to spatial mode 11. The photons from spatial modes 10 and 11 then pass through HWPs$^{22.5^\circ}$ and are measured in the $\{|D\rangle, |A\rangle\}$ basis by single-photon detectors D$_1$ and D$_2$, respectively.
Meanwhile, the photons from spatial modes 6 and 8 are directly routed to PBSs and measured in the $\{|H\rangle, |V\rangle\}$ basis by single-photon  detectors D$_3$, D$_4$, D$_5$, and D$_6$. These optical elements collectively evolve the state as
\begin{eqnarray}
\begin{split}             \label{eq18.7}
|\Phi'_7\rangle=&\;\frac{1}{2\sqrt{2}}[\alpha_1\alpha_2\alpha_3(|D\rangle_{10}+|A\rangle_{11})|H\rangle_5(|H\rangle_{12}|H\rangle_7+|V\rangle_{13}|V\rangle_7)
                  (|H\rangle_{14}|H\rangle_9+|V\rangle_{15}|V\rangle_9)\\&\;
                 %%%%%
                 +\alpha_1\alpha_2\beta_3(|D\rangle_{10}+|A\rangle_{11})|H\rangle_5(|H\rangle_{12}|H\rangle_7+|V\rangle_{13}|V\rangle_7)
                  (|H\rangle_{14}|V\rangle_9+|V\rangle_{15}|H\rangle_9)\\&\;
                 %%%%%
                 +\alpha_1\beta_2\alpha_3(|D\rangle_{10}+|A\rangle_{11})|H\rangle_5(|H\rangle_{12}|V\rangle_7+|V\rangle_{13}|H\rangle_7)
                  (|H\rangle_{14}|H\rangle_9+|V\rangle_{15}|V\rangle_9)\\&\;
                 %%%%%
                 +\alpha_1\beta_2\beta_3(|D\rangle_{10}+|A\rangle_{11})|H\rangle_5(|H\rangle_{12}|V\rangle_7+|V\rangle_{13}|H\rangle_7)
                  (|H\rangle_{14}|V\rangle_9+|V\rangle_{15}|H\rangle_9)\\&\;
                 %%%%%
                 +\beta_1\alpha_2\alpha_3(|D\rangle_{10}-|A\rangle_{11})|V\rangle_5(|H\rangle_{12}|V\rangle_7+|V\rangle_{13}|H\rangle_7)
                  (|H\rangle_{14}|V\rangle_9+|V\rangle_{15}|H\rangle_9)\\&\;
                 %%%%%
                 +\beta_1\alpha_2\beta_3(|D\rangle_{10}-|A\rangle_{11})|V\rangle_5(|H\rangle_{12}|V\rangle_7+|V\rangle_{13}|H\rangle_7)
                  (|H\rangle_{14}|H\rangle_9+|V\rangle_{15}|V\rangle_9)\\&\;
                 %%%%%
                 +\beta_1\beta_2\alpha_3(|D\rangle_{10}-|A\rangle_{11})|V\rangle_5(|H\rangle_{12}|H\rangle_7+|V\rangle_{13}|V\rangle_7)
                  (|H\rangle_{14}|V\rangle_9+|V\rangle_{15}|H\rangle_9)\\&\;
                 %%%%%
                 +\beta_1\beta_2\beta_3(|D\rangle_{10}-|A\rangle_{11})|V\rangle_5(|H\rangle_{12}|H\rangle_7+|V\rangle_{13}|V\rangle_7)
                  (|H\rangle_{14}|H\rangle_9+|V\rangle_{15}|V\rangle_9)].
\end{split}
\end{eqnarray}
We expand Eq.~\eqref{eq18.7} to obtain an intuitive representation of $|\Phi'_7\rangle$, which is given by
\begin{eqnarray}
\begin{split}             \label{eq19}
|\Phi_7'\rangle =&\;\frac{1}{2\sqrt{2}}\big[
      I_5I_7I_9
      C_{579}^3|\phi'\rangle_{579}\otimes|D\rangle_{10}|H\rangle_{12}|H\rangle_{14}+
      I_5I_7X_9
      C_{579}^3|\phi'\rangle_{579}\otimes|D\rangle_{10}|H\rangle_{12}|V\rangle_{15}\\&\;+
      I_5X_7I_9
      C_{579}^3|\phi'\rangle_{579}\otimes|D\rangle_{10}|V\rangle_{13}|H\rangle_{14}+
      I_5X_7X_9
      C_{579}^3|\phi'\rangle_{579}\otimes|D\rangle_{10}|V\rangle_{13}|V\rangle_{15}\\&\;+
      Z_5I_7I_9
      C_{579}^3|\phi'\rangle_{579}\otimes|A\rangle_{11}|H\rangle_{12}|H\rangle_{14}+
      Z_5I_7X_9
      C_{579}^3|\phi'\rangle_{579}\otimes|A\rangle_{11}|H\rangle_{12}|V\rangle_{15}\\&\;+
      Z_5X_7I_9
      C_{579}^3|\phi'\rangle_{579}\otimes|A\rangle_{11}|V\rangle_{13}|H\rangle_{14}+
      Z_5X_7X_9
      C_{579}^3|\phi'\rangle_{579}\otimes|A\rangle_{11}|V\rangle_{13}|V\rangle_{15}
\big].
\end{split}
\end{eqnarray}
Here the state
\begin{eqnarray}
\begin{split}             \label{eq20}
|\phi'\rangle_{579}=&\;(\alpha_1|H\rangle_5+\beta_1|V\rangle_5)
                     \otimes(\alpha_2|H\rangle_7+\beta_2|V\rangle_7)
                     \otimes(\alpha_3|H\rangle_9+\beta_3|V\rangle_9),
\end{split}
\end{eqnarray}
has the same form as the initial photon state in equations \eqref{eq13}-\eqref{eq15}.

\begin{table} [!h] \renewcommand{\arraystretch}{1.1}  % 增大行间距
\centering \caption{\justifying The correspondence between the clicks of the single-photon detector and feed-forward operations for implementing the three-photon $C^3$ gate in spatial modes 5, 7, and 9.}
\begin{tabular}{cccccccccccccc}%{lp{1.2in} lp{1.2in}}
\hline

\multicolumn {3}{c}{Measurement}  &     & \multicolumn {3}{c}{Feed-forward}     \\ \cline{1-3}  \cline{5-7}

    Alice   &  \;\;\;      Bob    &   \;\;\;    Charlie      & \;\,      &  \;\;\;    Mode 5   &  \;\;\;      Mode 7    &  \;\;\;     Mode 9                   \\

\hline

D$_1$  &\;\;\;  D$_3$  &\;\;\; D$_5$   &\;\;\;     &\;\;\; --    &\;\;\;  --       &\;\;\; --                    \\

D$_1$  &\;\;\;  D$_3$  &\;\;\; D$_6$   &\;\;\;     &\;\;\;  --   &\;\;\;  --       &\;\;\; $X$                    \\

D$_1$  &\;\;\;  D$_4$  &\;\;\; D$_5$   &\;\;\;     &\;\;\;  --   &\;\;\;  $X$      &\;\;\; --                    \\

D$_1$  &\;\;\;  D$_4$  &\;\;\; D$_6$   &\;\;\;     &\;\;\;  --   &\;\;\;  $X$      &\;\;\; $X$                    \\

D$_2$  &\;\;\;  D$_3$  &\;\;\; D$_5$   &\;\;\;     &\;\;\;  $Z$   &\;\;\;  --      &\;\;\; --                    \\

D$_2$  &\;\;\;  D$_3$  &\;\;\; D$_6$   &\;\;\;     &\;\;\;  $Z$   &\;\;\;  --      &\;\;\; $X$                    \\

D$_2$  &\;\;\;  D$_4$  &\;\;\; D$_5$   &\;\;\;     &\;\;\;  $Z$   &\;\;\;  $X$     &\;\;\; --                    \\

D$_2$  & \quad  D$_4$  & \quad D$_6$   & \quad     & \quad  $Z$   &\;\;\;  $X$     &\;\;\; $X$                    \\
\hline
\end{tabular}\label{Table2}
\end{table}

Finally, three-photon coincidence detection is performed among Alice, Bob, and Charlie. Based on equation \eqref{eq19}, the corresponding detector response patterns and required feed-forward operations for completing the $C^3$ gate are summarized in table \ref{Table2}. For instance, clicks of detectors D$_1$, D$_3$, and D$_5$ directly implement the $C^3$ gate in spatial modes 5, 7, and 9, whereas clicks of D$_1$, D$_3$, and D$_6$ require an additional feed-forward $X$ operation on spatial mode 9. The $X$ and $Z$ feed-forward operations in  table \ref{Table2} can be realized by using an HWP set at $45^\circ$ and $0^\circ$, respectively.

Putting all the pieces  together, one can see that the optical setup in figure~\ref{optical3} implements a three-photon $C^3$ gate with a success probability of $\frac{1}{2} \times \frac{1}{4} = \frac{1}{8}$  by assisting a three-photon GHZ state. The protocol succeeds if and only if each output mode (4--9) contains exactly one photon, otherwise, the scheme fails. The success of the scheme is heralded by three-fold coincidences, corresponding to the detection of exactly one photon at each of the three parties.

\subsection{Heralded long-distance $C^n$ gate}  \label{Sec3.2}

The optical approach shown in figure \ref{optical3} can be extended to implement a heralded and nondestructive $n$-qubit fan-out $C^n$ gate  acting on $n$ photonic qubits that are distributed among $n$ spatially separated parties, Alice, Bob, $\ldots$, and Nancy.

\begin{figure}   [tpb]%[!h]
\begin{center}
\includegraphics[width=8.3 cm,angle=0]{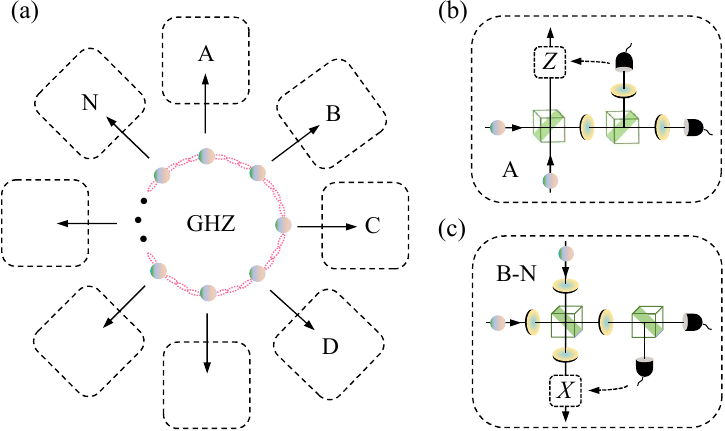}
\caption{\justifying Schematic optical setup to implement heralded and nondestructive long-distance $C^n$ gate among $n$ spatially separated parties.
(a) An $n$-photon GHZ state is distributed among the $n$ participants as an ancillary  entangled resource.
(b) Local optical module employed by Alice to manipulate her two photons $A$ and $a$. (c) Identical optical modules used by the remaining parties (Bob to Nancy) to manipulate their respective photon pairs. The letters $A$ to $N$ denote the parties from Alice to Nancy.}
\label{opticalN}
\end{center}
\end{figure}

A schematic diagram of the proposed setup is shown in figure \ref{opticalN}.
In this protocol, each party initially holds a single photon prepared in an arbitrary polarization state $\alpha_i |H\rangle + \beta_i|V\rangle$.
Meanwhile, an $n$-photon GHZ state $|\text{GHZ}\rangle_{a,b, \dots, n}$ as an ancillary entangled resource is distributed among the $n$ parties (see figure \ref{opticalN}(a)).
The initial state of the composite system is given by
\begin{eqnarray}
\begin{split}             \label{eq16}
|\Psi'_0\rangle = &\;(\alpha_1 |H\rangle_A + \beta_1|V\rangle_A) \otimes
                  (\alpha_2 |H\rangle_B + \beta_2|V\rangle_B)  \otimes \ldots \otimes (\alpha_n |H\rangle_N+\beta_n|V\rangle_N)
  \otimes |\text{GHZ}\rangle_{a,b, \dots, n}
\\= &\;\frac{1}{\sqrt{2}}(c_1|H\rangle^{\otimes n}+c_2|H\rangle^{\otimes (n-1)}|V\rangle+ \ldots +c_{2^n}|V\rangle^{\otimes n})(|H\rangle^{\otimes n} + |V\rangle^{\otimes n})_{a,b, \dots, n}.
\end{split}
\end{eqnarray}
At this point, each party holds two photons, one from the initial input state prepared by themselves and another one from the ancillary GHZ state.

A similar arrangement as that made in section \ref{Sec3.1}, Alice leads her two photons to the local optical module shown in figure \ref{opticalN}(b), while Bob, Charlie, $\ldots$, and Nancy respectively lead their individual photon pairs to identical modules illustrated in figure \ref{opticalN}(c).
In all modules, by post-selecting the event in which each of the spatial modes contains exactly one photon.
Then $|\Psi'_0\rangle$ will collapse to
\begin{eqnarray}
\begin{split}             \label{eq29}
|\Psi'_1\rangle =&\;\frac{1}{\sqrt{2^n}} \big[
I_A U_B \dots U_N C^n |\omega\rangle_{AB\dots N} \otimes |D\rangle_a
\sum |\vec{\nu}\rangle \\
&+ Z_A U_B \dots U_N C^n |\omega\rangle_{AB\dots N} \otimes |A\rangle_a
\sum |\vec{\nu}\rangle
\big],
\end{split}
\end{eqnarray}
which corresponds to the output state shown in equation \eqref{eq11}.
Here $|\omega\rangle_{AB\dots N}=c_1|H\rangle^{\otimes n}+c_2|H\rangle^{\otimes (n-1)}|V\rangle+ \ldots +c_{2^n}|V\rangle^{\otimes n}$, and $|\vec{\nu}\rangle = |\nu_b\rangle |\nu_c\rangle \dots |\nu_n\rangle$ denotes the joint polarization state of the $(n-1)$ photons held by Bob to Nancy, with $\nu_k \in \{H, V\}$. The summations are taken over all $2^{n-1}$ possible polarization state configurations.

After photon $a$ is measured in the $\{|D\rangle, |A\rangle\}$ basis, and photons $b$, $c$, $\ldots$, $n$ are measured in the $\{|H\rangle, |V\rangle\}$ basis, the detector-dependent classical feed-forward operations $U_k$ by the $k$th party ($k = B, \dots, N$) are applied to photons $A$, $B$, $\ldots$, $N$ to complete the protocol. The feed-forward operations are specified as follows:

(i) If the measurement outcomes are $|D\rangle_a$ and $|H\rangle_k$, no correction is required, i.e., $U_k = I_k$.

(ii) If the measurement outcomes are $|D\rangle_a$ and $|V\rangle_k$, the correction $U_k = X_k$ is applied on the $k$th party.

(iii) If the measurement outcomes are $|A\rangle_a$ and $|H\rangle_k$, a Pauli $Z$ operation is applied on Alice.

(iv) If the measurement outcomes are $|A\rangle_a$ and $|V\rangle_k$, Alice applies $Z$ operation and the $k$th party applies $U_k = X_k$.

The success of the scheme is heralded by the detection of exactly one single photon at each party. Under this heralding condition, the remaining output photons--one at each party's site--undergo the desired long-distance $C^n$ gate operation.
The success probability of the gate is $\frac{1}{2^n}$ and the protocol requires an
$n$-photon GHZ state.

\subsection{Post-selected long-distance $C^n$ gate}  \label{Sec3.3}

The post-selected approach is an alternative implementation of optical quantum gates, in which successful operation is confirmed by measuring ancillary and qubit modes.
In this subsection, we first show that the post-selected long-distance  $C^3$
gate can be implemented using the ancillary state generating from one Bell pair and one single photon.

\begin{figure}   [tpb]%[!h]
\begin{center}
\includegraphics[width=6.3 cm,angle=0]{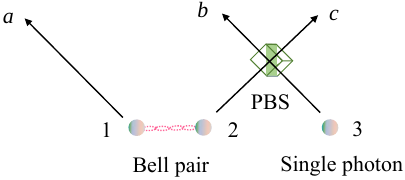}
\caption{\justifying The generation of an ancillary entangled state $|\psi\rangle_{abc}=\frac{1}{2}(
 |H\rangle_a |H\rangle_b|H\rangle_c
+|V\rangle_a |V\rangle_b|V\rangle_c
+|H\rangle_a |H\rangle_c|V\rangle_c
+|V\rangle_a |V\rangle_b|H\rangle_b)$ for implementing the post-selected long-distance  $C^3$ gate.}
\label{generation1}
\end{center}
\end{figure}

The protocol begins with the generation of an ancillary entangled state.
As shown in figure~\ref{generation1}, one polarization Bell pair and one single photon are initially prepared as
\begin{eqnarray}
\begin{split}             \label{eq123}
|\psi\rangle_0=&\;\frac{1}{2} (|H\rangle_1|H\rangle_2+|V\rangle_1|V\rangle_2)
\otimes (|H\rangle_3+|V\rangle_3).
\end{split}
\end{eqnarray}
After photons 2 and 3 pass through a PBS in figure~\ref{generation1}, the resulting three-photon state is given by
\begin{eqnarray}
\begin{split}             \label{eq124}
|\psi\rangle_{abc}=&\;\frac{1}{2}
(|H\rangle_a |H\rangle_b|H\rangle_c
+|V\rangle_a |V\rangle_b|V\rangle_c
+|H\rangle_a |H\rangle_c|V\rangle_c
+|V\rangle_a |V\rangle_b|H\rangle_b).
\end{split}
\end{eqnarray}
The first two terms of $|\psi\rangle_{abc}$ form a prefect GHZ state, which has been shown as an ancillary state in section~\ref{Sec3.1} to enable the implementation of a heralded  $C^3$ gate.
We next demonstrate that the state $|\psi\rangle_{abc}$ can also serve as an ancillary state to implement a post-selected $C^3$ gate. To this end, it suffices to show that the last two terms of $|\psi\rangle_{abc}$ remain unaffected by the implementation of the $C^3$ gate.

In fact, the last two terms of equation~\eqref{eq124} can be rewritten as
\begin{eqnarray}
\begin{split}             \label{A1}
|\psi'\rangle_{abc}=\frac{1}{\sqrt{2}}(|H\rangle_a |H\rangle_c|V\rangle_c
+|V\rangle_a |V\rangle_b|H\rangle_b).
\end{split}
\end{eqnarray}
When this state is used as the ancillary state in the setup shown in figure~\ref{optical3}, the initial state of the entire system can be expressed as
\begin{eqnarray}
\begin{split}             \label{A2}
|\varphi_0\rangle=|\psi'\rangle_{abc}\otimes|\phi'\rangle_{A}\otimes|\phi'\rangle_{B}\otimes|\phi'\rangle_{C}.
\end{split}
\end{eqnarray}

Firstly, photons $A$ and $a$ first pass through PBS$_1$ in figure~\ref{optical3}.
By postselecting the events in which output modes 4 and 5 each contain exactly one photon, the state $|\varphi_0\rangle$ is collapsed into
\begin{eqnarray}
\begin{split}             \label{A3}
|\varphi_1\rangle=&\;\frac{1}{\sqrt{2}}(\alpha_1 |H\rangle_4 |H\rangle_5 |H\rangle_3 |V\rangle_3
                   + \beta_1 |V\rangle_5 |V\rangle_4 |V\rangle_2 |H\rangle_2)
             \otimes(\alpha_2|H\rangle_{B} + \beta_2|V\rangle_{B}) \\&
             \otimes(\alpha_3|H\rangle_{C} + \beta_3|V\rangle_{C}).
\end{split}
\end{eqnarray}

\begin{figure*}   [tpb]%[!h]
\begin{center}
\includegraphics[width=14 cm,angle=0]{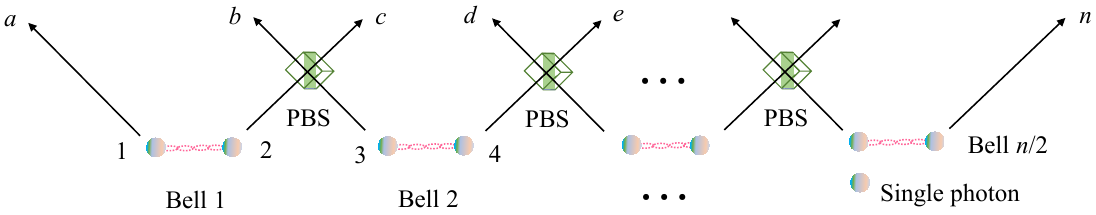}
\caption{\justifying The generation of an $n$-qubit ancillary entangled state $|\psi\rangle_{ab\ldots n}$ for implementing the post-selected long-distance $C^n$ gate. When $n$ is even, $\frac{n}{2}$ Bell pairs are required; when $n$ is odd, $\frac{n-1}{2}$ Bell pairs and one single photon are required.}
\label{generation2}
\end{center}
\end{figure*}

Subsequently, photons $B$, $C$, and the photons emitted from modes 2 and 3, pass through four HWPs set at $22.5^\circ$ and the PBS$_2$ and PBS$_3$.
The state $|\varphi_1\rangle$ then evolves as
\begin{eqnarray}
\begin{split}             \label{A4}
|\varphi_2\rangle=&\;\frac{1}{4}[\alpha_1\alpha_2\alpha_3 |H\rangle_4 |H\rangle_5 (|H\rangle_8 |H\rangle_8-|V\rangle_9 |V\rangle_9)
(|H\rangle_7 |H\rangle_9+|V\rangle_6 |H\rangle_9+|H\rangle_7 |V\rangle_8
+|V\rangle_6 |V\rangle_8)\\&
+\alpha_1\alpha_2\beta_3 |H\rangle_4 |H\rangle_5 (|H\rangle_8 |H\rangle_8
-|V\rangle_9 |V\rangle_9)
(|H\rangle_7 |H\rangle_9+|V\rangle_6 |H\rangle_9
-|H\rangle_7 |V\rangle_8 -|V\rangle_6 |V\rangle_8)\\&
+\alpha_1\beta_2\alpha_3 |H\rangle_4 |H\rangle_5
(|H\rangle_8 |H\rangle_8 -|V\rangle_9 |V\rangle_9)
(|H\rangle_7 |H\rangle_9
-|V\rangle_6 |H\rangle_9+|H\rangle_7 |V\rangle_8 -|V\rangle_6 |V\rangle_8) \\&
+\alpha_1\beta_2\beta_3 |H\rangle_4 |H\rangle_5 (|H\rangle_8 |H\rangle_8-|V\rangle_9 |V\rangle_9) (|H\rangle_7 |H\rangle_9-|V\rangle_6 |H\rangle_9-|H\rangle_7 |V\rangle_8
+|V\rangle_6 |V\rangle_8) \\&
+\beta_1\alpha_2\alpha_3 |V\rangle_4 |V\rangle_5 (|H\rangle_6 |H\rangle_6
-|V\rangle_7 |V\rangle_7)
                    (|H\rangle_7 |H\rangle_9+|V\rangle_6 |H\rangle_9
                    +|H\rangle_7 |V\rangle_8
                    +|V\rangle_6 |V\rangle_8) \\&
+\beta_1\alpha_2\beta_3 |V\rangle_4 |V\rangle_5 (|H\rangle_6 |H\rangle_6
-|V\rangle_7|V\rangle_7)
(|H\rangle_7 |H\rangle_9  +|V\rangle_6 |H\rangle_9
-|H\rangle_7 |V\rangle_8 -|V\rangle_6 |V\rangle_8)  \\&
+\beta_1\beta_2\alpha_3 |V\rangle_4 |V\rangle_5
 (|H\rangle_6 |H\rangle_6 -|V\rangle_7 |V\rangle_7)
 (|H\rangle_7 |H\rangle_9
 -|V\rangle_6 |H\rangle_9+|H\rangle_7 |V\rangle_8-|V\rangle_6 |V\rangle_8) \\&
+\beta_1\beta_2\beta_3 |V\rangle_4 |V\rangle_5
(|H\rangle_6 |H\rangle_6-|V\rangle_7 |V\rangle_7)
(|H\rangle_7 |H\rangle_9
-|V\rangle_6 |H\rangle_9-|H\rangle_7 |V\rangle_8
+|V\rangle_6 |V\rangle_8)].
\end{split}
\end{eqnarray}

Based on equation~\eqref{A4}, one can see that two photons occupy one of the output modes (6, 7, 8, or 9) in each term.
Hence, it becomes impossible for each of the output modes 4–9 to contain exactly one photon simultaneously.
Therefore, when the ancillary source is $|\psi\rangle_{abc}$, the protocol presented in figure~\ref{optical3} can successfully implement the long-distance $C^3$ gate if and only if each of the output modes 4–9 contains exactly one photon.
The overall success probability of the post-selected $C^3$ gate is $\frac{1}{8}$, and one Bell pair and one single photon are required.

Similarly, as shown in figure \ref{generation2}, such an $n$-qubit ancillary entangled state $|\psi\rangle_{ab\ldots n}$ can be deterministically generated from $\frac{n}{2}$ Bell pairs for even $n$ or $\frac{n-1}{2}$ Bell pairs and one single photon for odd $n$ by using multiple PBSs ~\cite{pan2001experimental,lu2007experimental,zhong201812}.
This ancillary state contains an $n$-qubit GHZ component corresponding to the events in which each mode contains exactly one photon, together with additional components in which multiple photons occupy the same mode.
As discussed in the above analysis, the presence of these additional components does not affect the result of the $C^n$ gate when each output mode in figure~\ref{opticalN} contains exactly one photon.

\section{Discussions}  \label{Sec4}

\subsection{The resource cost and the success probability over previous protocols}  \label{Sec4.1}

The difference between the heralded and post-selected schemes lies in the used ancillary states. Specifically, the heralded scheme requires a heralded $n$-photon polarization-entangled GHZ state. Several methods have been proposed to generate such states, including entanglement swapping of multiple entangled pairs~\cite{lu2009experimental} and single-photon interference~\cite{GHZ2008,GHZ2024,cao2024photonic,chen2024heralded,forbes2025heralded}. In general, generating a heralded $n$-photon GHZ state requires $2n$ single photons and succeeds with probability $(\frac{1}{2})^{2n-1}$ \cite{GHZ2024}.
Therefore, the total success probability of our heralded protocol is $(\frac{1}{2})^{2n-1}\times (\frac{1}{2})^{n}=(\frac{1}{2})^{3n-1}$ and $2n$ ancillary single photons are required. Similarly, heralded Bell pairs can be generated using four single photons with a success probability of $\frac{1}{8}$ ~\cite{GHZ2024}.
Previous studies have demonstrated that a heralded photonic CNOT gate can be implemented using either one ancillary Bell pair or two single photons, with success probabilities of
$\frac{1}{4}$~\cite{pittman2001probabilistic,zeuner2018integrated,zhao2005experimental} and
$\frac{1}{8}$~\cite{li2021heralded}, respectively. Based on these results and the fact that the decomposition-based $C^n$ gate requires $(2n-2)$ CNOT gates and $(n-1)$ Bell pairs, the decomposition-based implementation of the $C^n$ gate requires $12(n-1)$ ancillary single photons with a success probability of $\left(\frac{1}{8}\right)^{3n-3}\times\left(\frac{1}{4}\right)^{2n-2}=\left(\frac{1}{2}\right)^{13n-13}$ \cite{pittman2001probabilistic,zeuner2018integrated,zhao2005experimental} or $8(n-1)$ ancillary single photons with a success probability of $\left(\frac{1}{8}\right)^{3n-3}$ \cite{li2021heralded}.

For our post-selected scheme, the ancillary $n$-photon state can be generated by using $\frac{n}{2}$ Bell pairs for even $n$, or $\frac{n-1}{2}$ Bell pairs and one single photon for odd $n$~\cite{pan2001experimental,lu2007experimental,zhong201812} (see figure~\ref{generation2}).
Therefore, for even $n$,
the overall success probability of our post-selected $C^n$ gate is $\left(\frac{1}{8}\right)^{\frac{n}{2}}\times\left(\frac{1}{2}\right)^{n}=\left(\frac{1}{2}\right)^{\frac{5n}{2}}$ and $2n$ ancillary single photons are required. For odd $n$, the overall success probability is $\left(\frac{1}{8}\right)^{\frac{n-1}{2}}\times\left(\frac{1}{2}\right)^{n}=\left(\frac{1}{2}\right)^{\frac{5n-3}{2}}$ and $(2n-1)$ ancillary single photons are required.
In contrast,  the success probability of a post-selected $C^n$ gate based on the decomposition-based circuit is $(\frac{1}{8})^{n-1}\times(\frac{1}{9})^{2n-2}=(\frac{1}{648})^{n-1}$ and $4(n-1)$ ancillary single photons are required.
This is because the optimal success probability of a post-selected optical CNOT gate without any ancillary photons is $\frac{1}{9}$~\cite{okamoto2005demonstration}. The comparisons between our protocols and the previous ones in terms of required ancillary single photons and the overall success probability are presented in table \ref{Table4}.

\begin{table*} [tb] \renewcommand{\arraystretch}{1.2}  % 增大行间距
\centering \caption{\justifying Comparisons between our long-distance $C^n$ gate and the previous counterparts in terms of ancillary single photons and success probability.}
\begin{tabular}{cccccccccccccc}%{lp{1.2in} lp{1.2in}}
\hline

    Schemes    & \;   Ancillary single photons        & \;    Probability     \\

\hline

Heralded decomposition \cite{baumer2025measurement,pittman2001probabilistic,zeuner2018integrated,zhao2005experimental}     & \; $12(n-1)$        & \;  $(\frac{1}{2})^{13n-13}$                    \\

Heralded decomposition \cite{baumer2025measurement,li2021heralded}    & \; $8(n-1)$       & \;  $(\frac{1}{8})^{3n-3}$                   \\

Post-selected decomposition \cite{baumer2025measurement,okamoto2005demonstration}    & \; $4(n-1)$       & \;  $(\frac{1}{648})^{n-1}$ \\

Heralded ours                       & \; $2n$       & \;  $(\frac{1}{2})^{3n-1}$                  \\

Post-selected  ours                                        & \;  $2n$ or $2n-1$      & \; $(\frac{1}{2})^{\frac{5n}{2}}$  or  $(\frac{1}{2})^{\frac{5n-3}{2}}$                    \\

                             \hline
\end{tabular}\label{Table4}
\end{table*}

\subsection{The effect of depolarizing noise on the protocol}  \label{Sec4.2}

Our scheme requires a pre-shared $n$-qubit GHZ state to implement the long-distance $C^n$ gate. However, in practice, the $n$-qubit GHZ state inevitably suffers from decoherence during long-distance distribution. Depolarizing noise is the main source of such decoherence. In the following, we evaluate the impact of depolarizing noise on our proposed scheme.

An initial maximally entangled $n$-qubit GHZ state affected by depolarizing noise can be modeled as a generalized Werner state \cite{eltschka2012entanglement}
\begin{eqnarray}
\begin{split}             \label{eq23}
\mathcal{D}_p(\rho_{\text{GHZ}_1}^{ab\dots n}) =&\;(1-p)\rho_{\text{GHZ}_1}^{ab\dots n}+\frac{p}{2^n}\mathbb{I},
\end{split}
\end{eqnarray}
where $p\in[0,1]$ denotes the depolarizing probability, $\rho_{\text{GHZ}_1}^{ab\dots n}=|\psi_1\rangle_{ab\dots n}\langle\psi_1|$ is the ideal $n$-qubit GHZ state with $|\psi_1\rangle_{ab\dots n} = (|0\rangle^{\otimes n} + |1\rangle^{\otimes n})/\sqrt{2}$, and $\mathbb{I}$ is the identity operator on the $2^n$-dimensional Hilbert space.
Since the identity operator can be decomposed as a mixture of all $2^n$ orthogonal GHZ basis states $|\psi_k\rangle$, i.e., $\mathbb{I}=\sum_{k=1}^{2^n}|\psi_k\rangle\langle\psi_k|$, the equation~\eqref{eq23} can be rewritten as
\begin{eqnarray}
\begin{split}             \label{eq24}
\mathcal{D}_p(\rho_{\text{GHZ}_1}^{ab\dots n}) =&\;(1-p+\frac{p}{2^n})\rho_{\text{GHZ}_1}^{ab\dots n}+\frac{p}{2^n}\sum_{k=2}^{2^n}|\psi_k\rangle\langle\psi_k|.
\end{split}
\end{eqnarray}
When $p=0$, it means our protocol faithfully implements the $n$-qubit fan-out gate $C^n$ in the ideal scenario, and it can be expressed by the following density operator
\begin{eqnarray}
\begin{split}             \label{eq25}
\sum_{m=1}^{2^n}V_mM_mU(\rho_\text{in}\otimes\rho_{\text{GHZ}_1}^{ab\dots n})U^\dagger M_m^\dagger V_m^\dagger=|\Psi_\text{id}\rangle\langle\Psi_\text{id}|.
\end{split}
\end{eqnarray}
Here $\rho_\text{in}=|\phi\rangle_{AB\dots N}\langle\phi|$ is the density operator of initial $n$-qubit state in equation \eqref{eq8}.
$U$ denotes the collective unitary operations on the state $\rho_\text{in}\otimes\rho_{\text{GHZ}_1}^{ab\dots n}$.
$M_m$ are the measurement operators associated with outcome $|m\rangle$, and $V_m$ are the corresponding feed-forward operations.
$|\Psi_\text{id}\rangle=C^n|\phi\rangle_{AB\dots N}$ is the ideal output state.

Note that any single GHZ state can serve as the ancillary channel to faithfully implement the $C^n$ gate. However, when the ancillary channel becomes a mixture of multiple orthogonal GHZ basis states, the fidelity of the $C^n$ gate will be decreased.
After the ancillary GHZ state is affected by depolarizing noise, the actual output state of the overall system is
\begin{eqnarray}
\begin{split}             \label{eq26}
\rho_\text{out}&= \sum_{m=1}^{2^n}V_mM_mU[\rho_\text{in}\otimes\mathcal{D}_p(\rho_{\text{GHZ}_1}^{ab\dots n})]U^\dagger M_m^\dagger V_m^\dagger\\&\;
=(1-p+\frac{p}{2^n})|\Psi_\text{id}\rangle\langle\Psi_\text{id}|+(p-\frac{p}{2^n})|\Psi_\text{nid}\rangle\langle\Psi_\text{nid}|.
\end{split}
\end{eqnarray}
Here the state $|\Psi_\text{nid}\rangle\langle\Psi_\text{nid}|=\sum_{m=2}^{2^n}V_mM_mU(\rho_\text{in}\otimes \sum_{k=2}^{2^n}|\psi_k\rangle\langle\psi_k|)U^\dagger M_m^\dagger V_m^\dagger$ represents the nonideal contribution to the output state that arises from the mixture of the remaining ($2^n-1$) undesired ancillary GHZ states.

\begin{figure}   [tpb]%[!h]
\begin{center}
\includegraphics[width=8 cm,angle=0]{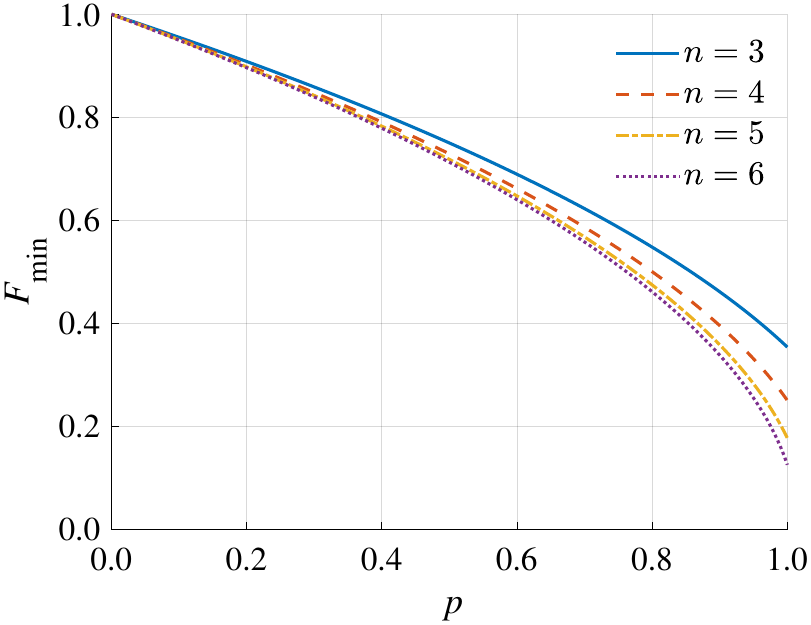}
\caption{\justifying Lower bound of the fidelity of the $n$-qubit fan-out gate $C^n$, $F_{\min} = \sqrt{1 - p + \tfrac{p}{2^n}}$, as a function of the depolarizing probability $p$ and the number of qubit $n$.
}
\label{fidelity}
\end{center}
\end{figure}

The performance of the gate can be quantified by the fidelity~\cite{nielsen2010quantum}
\begin{eqnarray}
\begin{split}             \label{eq27}
F= \sqrt{\langle\Psi_\text{id}|\rho_\text{out}|\Psi_\text{id}\rangle},
\end{split}
\end{eqnarray}
which evaluates the overlap between the actual and the ideal output states. Substituting equation~\eqref{eq26} into equation~\eqref{eq27} yields
\begin{eqnarray}
\begin{split}             \label{eq28}
F = \sqrt{1-p+\frac{p}{2^n}+(p-\frac{p}{2^n})|\langle\Psi_\text{id}|\Psi_\text{nid}\rangle|^2}  \geq \sqrt{1-p+\frac{p}{2^n}}.
\end{split}
\end{eqnarray}

This result indicates that the gate fidelity retains a lower bound of $F_\text{min}=\sqrt{1-p+\frac{p}{2^n}}$ in the presence of depolarizing noise.
Figure \ref{fidelity} illustrates the relation between $F_\text{min}$ and the number of qubit $n$ and the depolarizing probability $p$. When the depolarizing error $p$ is small, high fidelity can be achieved. If applying the entanglement purification techniques to the distributed GHZ state, the noise parameter $p$ can be suppressed \cite{huber2011purification,yan2023advances,victora2023entanglement,shimizu2025simple}.

\subsection{The effect of photon loss on the protocol}  \label{Sec4.3}

Photon loss is a critical factor affecting the efficiency of our optical scheme.
The overall efficiency of our protocol arises from five contributions: (i) the intrinsic success probability $P_n$ of the $C^n$ gate itself; (ii) the survival efficiency $\eta_{\mathrm{GHZ}}$ of the auxiliary $n$-photon GHZ state accounting for photon loss; (iii) the single-photon source efficiency $\eta_s$; (iv) the per-photon optical transmission efficiency $\eta_t$, which characterizes propagation losses introduced by PBSs and HWPs; and (v) the single-photon detector efficiency $\eta_d$. Combining these terms, the total efficiency of our $n$-qubit optical heralded $C^n$ gate is expressed as
\begin{eqnarray}
\begin{split}             \label{eq42}
\eta_{C^n} = P_n \times \eta_{\mathrm{GHZ}} \times \eta_s^n \times \eta_t^n \times \eta_d^n.
\end{split}
\end{eqnarray}

Specifically, the intrinsic success probability $P_n$ was discussed in Sec. \ref{Sec4.1}.
The per-photon survival efficiency for generating a three-photon GHZ state is given by $\frac{\eta_s}{2 - \eta_d \eta_s}$ \cite{GHZ2008}.
Adopting state-of-the-art experimental parameters $\eta_s=0.92$ \cite{photonSource,photonloss} and $\eta_t=\eta_d=0.98$ \cite{detector},
the total efficiency of the three-qubit optical heralded $C^3$ gate is
\begin{eqnarray}
\begin{split}             \label{eq43}
\eta_{C^3} = \left(\frac{1}{2}\right)^8 \times \left(\frac{0.92}{2-0.98\times 0.92}\right)^3 \times 0.92^3 \times 0.98^3  \times 0.98^3 = 1.6\times10^{-3}.
\end{split}
\end{eqnarray}

 \section{Conclusion}  \label{Sec5}

In conclusion, we have proposed a long-distance three-qubit fan-out gate and further generalized the approach to implement an $n$-qubit fan-out gate $C^n$ using an $n$-qubit GHZ state and $n$ local CNOT gates. Compared with the decomposition-based method \cite{baumer2025measurement}, our protocol reduces both the CNOT and Hadamard gate counts by ($n-2$).
Besides, we proposed two optical schemes for implementing heralded and post-selected $C^n$ gates, with success probabilities of $(\frac{1}{2})^{3n-1}$ and $(\frac{1}{2})^{\frac{5n}{2}}$ for even $n$ or $(\frac{1}{2})^{\frac{5n-3}{2}}$ for odd $n$, respectively. Both of these two schemes outperform the decomposition-based approaches \cite{baumer2025measurement}. These features render our scheme experimentally feasible with current photonic technologies and promising for scalable optical distributed quantum computation. We believe that the protocol can be extended to spatial-mode DOF \cite{Feng2024}, and other quantum hardware platforms including atom–cavity systems \cite{daiss2021quantum},  trapped ions \cite{main2024distributed}, and  superconducting qubits \cite{qiu2025deterministic}.

\data{The data that support the findings of this study are available upon reasonable request from the authors.}

\ack{This work was supported by the National Natural Science Foundation of China under Grant No. 12505028 and Grant No. 62371038, and  Science Research Project of Hebei Education Department under Grant No. QN2025054.}

\providecommand{\newblock}{}

\end{document}